\documentclass[showpacs,aps,pre,twocolumn,superscriptaddress,textsuperscript]{revtex4}

\usepackage{graphicx}
\usepackage{amssymb}
\usepackage[sort&compress]{natbib}

\begin{document}

\title{Resonance-induced acceleration of the RBNE-BNE segregation inversion of granular mixtures}

\author{Yufei Shao} \author{Anghao Li} \author{Zaizheng Wang} \author{Min Sun}
\author{Decai Huang}
\email[Corresponding author:] {hdc@njust.edu.cn}
\affiliation{Department of Applied Physics, Nanjing University of Science and Technology, Nanjing 210094, China}

\date{\today}

\begin{abstract}

  This paper presents the experiments and simulations on the resonance-induced acceleration of the reverse Brazil nut effect (RBNE)-Brazil nut effect (BNE) segregation inversion of binary mixtures in flat-bottom and circular-bottom containers. Both experimental and simulation results indicate that the starting location of the sinkage of heavier grains at the top layer is triggered with certain randomness in the flat-bottom container, whereas it first occurs at either of the lateral edges of the bottom in the circular-bottom container. The quantified segregation factors in simulations show that the transition from the RBNE segregation state to the BNE segregation state happens faster in the circular-bottom container than that in the flat-bottom container. The occurrence of standing-wave resonant spots of higher and lower granular temperature accelerates the RBNE-BNE segregation inversion. From the elastic collision model of single grain, the bottom with a larger angle leads to more energy transfer from the vertical direction to the horizontal direction. The theoretical predictions are confirmed by the simulations of a monodisperse granular bed. The flat-bottom container has a uniform distribution with a standing-wave period of granular temperature and packing density, whereas the circular-bottom container possesses a higher granular temperature in the horizontal direction and a lower packing density at the lateral edges of the circular bottom. Owing to the buoyancy effect, heavier grains easily sink first at the resonant spots with higher temperature.

\end{abstract}


\date{\today}

\maketitle
\section{Introduction}

Granular materials, consisting of substantial discrete grains, are among the most common materials in daily life and in nature\cite{Jaeger1996,Aranson2006,Guo2015FM47.21,Mark2018FM50.407,Durian2015NC6.6527}. Compared with the classical thermodynamic system, many properties of granular materials are unique due to their innate macroscopic size and dissipative interaction between the grains.
One special phenomenon is segregation, in which binary mixtures may spontaneously separate through the difference in their individual components, such as density, size, and surface roughness\cite{Schroter2006PRE74.011307,Lueptow2020PRF5.044301,
Shi2007PRE75.061302,Hsiau2020APT31.94,Lueptow2006PRE73.031304,Song2020PT372.40,Lueptow2020PRR2.022609}.
Owing to the intrinsic dissipation, external energy has to be continuously injected into granular system to keep grains in motion.
Various segregation agitation methods, i.e., shear and vibration, have been used to separate and mix binary grains\cite{Hill2011NJP13.095009,Yang2020AIChe67.17090,
Dong2022APT33.103551,Lueptow2021PRF6.054301,Sautel2021PRE103.022901}.
Many driving mechanism based on materials properties, i.e., percolation and buoyancy, have also been proposed to describe the segregation of binary mixtures\cite{Hong2001PRL86.3423,Breu2003PRL90.014302,Lueptow2019AR10.5,
Oshtorjani2021PRE103.062903,Wang2021SciAdv7.8737}.
Although great progress has been made, the underlying physical mechanism for the segregation remains interesting, and how to achieve a fast segregation of binary mixtures is expected to be deeply discovered and investigated for scientific relevance and practical operation.

When a granular system of binary mixtures with identical size is subjected to vertical vibration, mass density plays a key role in the segregation process.
The buoyancy effect is considered as a plausible driving mechanism, in which grains with higher mass density prefer to sink down first compared with lighter grains.
Brazil-nut effect (BNE) refers to a two-layer segregation pattern, in which heavier and lighter grains are at top and bottom layers, respectively.
On the contrary, reverse BNE (RBNE) is defined as the inverse segregation state\cite{Hong2001PRL86.3423,Breu2003PRL90.014302,HuertaPRL92.114301,
Schnautz2005PRL95.028001,Shi2007PRE75.061302,Du2011PRE84.041307}.
From a microscopic perspective, the buoyancy effect due to the difference in mass density is considered as the intrinsic driving force to decide the segregation tendency.
Meanwhile, recent investigations have shown that the injected external energy is not  distributed in the granular system, and results in a series of macroscopic collective behaviors, such as surface wave and pattern formation\cite{Huang2018PRE97.052905,Yule2014PRE89p022202,Zhao2022PRE105.L022902,
Zhao2022PRE105.L022902,Huang2011PRL107.028001,Sun2021POF33.123319,
Huang2012SM8.11939,Moon2003PRL91.134301,Melo1995PRL75.3838,Nature1996Umbanhowar382.793}.
Compared with the conventional thermodynamic system, the granular temperature defined by kinetic energy is always introduced to characterize these dynamic behaviors of the granular system.
For a vibrated granular bed in a container, the existence of a lateral wall strongly affects the spatial gradient in granular temperature and packing density.
For instance, a higher dissipative collision between the grain and the sidewall leads to an increase in packing density and granular temperature, in which the buoyancy effect leads to the sinkage of heavier and smaller grains along the sidewalls of the container\cite{Yule2013PRL111.038001,Hsiau2002AIChJ48.1430,Goldirsch1993PRL70.1619,
Martins2021PRE103.062901}.
With this macroscopic convection, lighter and larger grains are easily brought to the surface and settle there.

Recently, there have been a number of studies aiming to control the transition of the BNE and RBNE segregation states through the geometrical shape of the container.
Both experimental and simulation results show that a symmetrical sawtooth-shaped base leads to the vertical segregation of binary mixtures, similar to that observed in flat-bottom container\cite{Shi2009PRE80.061306,Shi2015PT286.629,
Mobarakabadi2017EPJE40.79,Bhateja2017PRF2.052301,Jing2017EPJ140.03056}.
However, an asymmetrical geometrical base induces a different spatial distribution of interfacial pressure gradients, which leads to a horizontal transport and segregation caused by density and size of the grains.
Loguinova and Windows-Yule et al. numerically analyzed the importance of the resonant effect on the energy transport and the segregation of binary mixtures in a vertically vibrated container\cite{Loguinova2009GM11.63,Yule2015NJP17.023015}.
They concluded that a granular bed has a natural resonant frequency according to grains' density and size.
This intrinsic characteristic was validated by subsequent experiments\cite{Cai2015PRE91.032204,Cai2020PRE101.032902}, in which continuous input energy led to a series of periodical resonant waves.
Furthermore, the saw-tooth bottom exerts a great influence on the induced horizontal flow of grains due to resonance.
Nevertheless, knowledge on how the nonuniform energy distribution due to the resonance effect and the bottom shape of the container exert influences on the segregation of binary mixtures is limited.

This study intends to compare the resonance-induced segregation of binary grains with different densities in flat-bottom and circular-bottom containers through experiments and simulations.
The external energy is input by vertically vibrating the container. Section \ref{SimMod} describes the experimental setup and simulation model.
Section \ref{SegPattern} shows the process of the RBNE-BNE segregation inversion and the temporal evolution of grain mass center for each species of grains by introducing a segregation factor.
Section \ref{SegMech} reveals the evolution of the temporal distribution of granular temperature for binary mixtures.
An elastic collision of single grain with a flat bottom and a circular bottom is proposed and corresponding validation by simulations of granular system of monodisperse grains.
Section \ref{Conclusions} concludes the study.
\begin{figure}[htbp]
\centering
\includegraphics[width=0.35\textwidth,trim=0 0 0 0,clip]{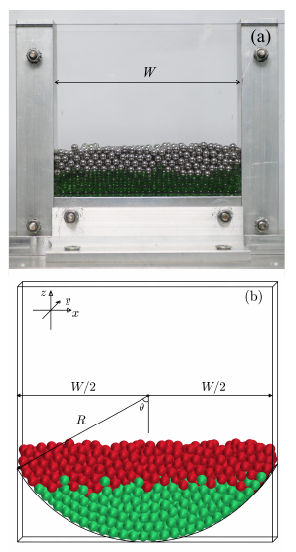}
  \caption{Snapshots of the system at the beginning. (a)Experimental setup with a flat bottom of $\theta=0$, (b)Simulation system with a circular bottom of $\theta=\pi/3$. The coordinate origin is placed at the center of the bottom for all experiments and simulations.}
\label{fig:FigModelExpSim}
\end{figure}
\section{Experimental setup and simulation model}
\label{SimMod}

Fig.\ref{fig:FigModelExpSim}(a) illustrates the experimental setup, which is a three-dimensional rectangular container. The frame is made of aluminum alloy with height $H=100~{\rm mm}$, width $W=100~{\rm mm}$, and thickness $T_c=20~{\rm mm}$. The front and rear walls are covered with two perspex plates.
The container is placed on a vertically vibrational shaker with a sinusoidal displacement in $z$ direction, $z_{\rm{vib}}=A{\rm{sin}(2{\pi}}ft)$, where $A$ and $f$ are the vibration amplitude and frequency, respectively.
The vibration period is $T=1/f$, and the maximum of vibration velocity is $v_0=2\pi Af$.
The same simulation system is used, as shown in Fig.\ref{fig:FigModelExpSim}(b), which has a circular bottom, $\theta=\pi/3$.
In the simulations, the bottom angle $\theta$ can be adjusted from $0$ to $\pi/2$, in which $\theta=0$ denotes a flat bottom with $R=\infty$, and $\theta=\pi/2$ is for a half circular bottom with $R=W/2$.
The maximum amplitude of dimensionless vibration acceleration is defined as ${\it{\Gamma}}=A(2{\pi}f)^2/g$, where $g=9.8~{\rm{m/s^2}}$ is the gravity acceleration.
In this work, $f$ and ${\it{\Gamma}}$ are set to $20~{\rm{Hz}}$ and $3.8$, respectively.

In the experiments, different mass densities of grains are realized by using various materials, stainless steel and glass, which have mass densities $\rho_h=7.8$ and $\rho_l=2.5~\rm{g/cm^3}$, respectively. $h$ and $l$ represent the heavier stainless steel and lighter glass grains, respectively.
The grains' radiuses are equal to $r_h=r_l=4.0\pm{0.02}~{\rm{mm}}$.
The segregation process can be influenced by the amount of grains and the number of layers in the container.
In this work, the numbers of lighter grains and heavy grains are set to $N_h=N_l=500$.

In the simulations, the grain motion is described using Newton's equations as in our previous works\cite{Huang2006PRE74.061301,Huang2012PRE85.031305,Huang2022PT401.117271}.
The acceleration due to gravity is set to $g=9.8~{\rm m/s^{2}}$, the same as that in the experiment.
The Verlet algorithm is used to update the positions and velocities of the grains at each simulation time step.
The interaction between two contacting grains is regarded as a soft model, in which the forces in the normal and tangential directions are considered. The normal interaction at the contact point is described by using the Cundall-Strack formula\cite{Cundall1979,Schafer1996}:
\begin{equation}
  F_{n}={\frac{4}{3}}E^{*}{\sqrt{R^{*}}}{\delta_{n}}^{3/2}
       -2{\sqrt{\frac{5}{6}}}{\beta}{\sqrt{S_{n}m^{*}}}V_{n}.
  \label{CundallFn}
\end{equation}

The tangential component is taken as the minor tangential force with a memory effect and the dynamic frictional force:
\begin{equation}
  F_{\tau}=-\min({S_{\tau} {\delta}_{\tau}}-2{\sqrt{\frac{5}{6}}}{\beta}{\sqrt{S_{\tau}m^{*}}}V_{\tau},{{\mu} F_{n}}){\rm{sign}}({\delta}_{\tau}).
  \label{CundallFt}
\end{equation}

In Eqs. (\ref{CundallFn}) and (\ref{CundallFt}), $n$ and $\tau$ respectively denote the normal and tangential directions at the contact point; and $\delta_{n}$ and $\delta_{\tau}$ respectively denote the normal and tangential displacements since time $t_0$ at which contact is first established. The calculation details are as follows:
\begin{equation}
  \beta={\frac{{\rm ln}e}{{\sqrt{{\rm ln}^{2}e+{\pi}^{2}}}}},~
  {\frac{1}{m^{*}}}={\frac{1}{m_{i}}}+{\frac{1}{m_{j}}}, \\  \nonumber
\end{equation}
\begin{equation}
  S_{n}=2E^{*}{\sqrt{R^{*}{\delta_{n}}}},~
  S_{\tau}=8E^{*}{\sqrt{R^{*}{\delta_{n}}}},\\  \nonumber
\end{equation}
\begin{equation}
{\frac{1}{E^{*}}}={\frac{1-{\nu_{i}^{2}}}{E_{i}}}+{\frac{1-{\nu_{j}^{2}}}{E_{j}}},~
{\frac{1}{R^{*}}}={\frac{1}{R_{i}}}+{\frac{1}{R_{j}}}, \\ \nonumber
\end{equation}

\noindent where $e$ is the coefficient of restitution. The quantities $m_i$ and $m_j$ are the masses of grains $i$ and $j$ making contact, respectively, and $S_{n}$ and $S_{\tau}$ characterize the normal and tangential stiffness of the grains, respectively. $E$ and ${\nu}$ are the Young's modulus and Poisson's ratio, respectively.
In our simulations, the friction coefficient $\mu$ is fixed for both heavier and lighter grains, and a collision between a grain and a wall is regarded as a grain-grain collision, except that the wall has infinite mass and diameter.
Table 1 lists the values of the material parameters of the grains in the simulations.
\begin{figure}[htbp]
\centering
\includegraphics[width=0.48\textwidth,trim=165 530 180 135,clip]{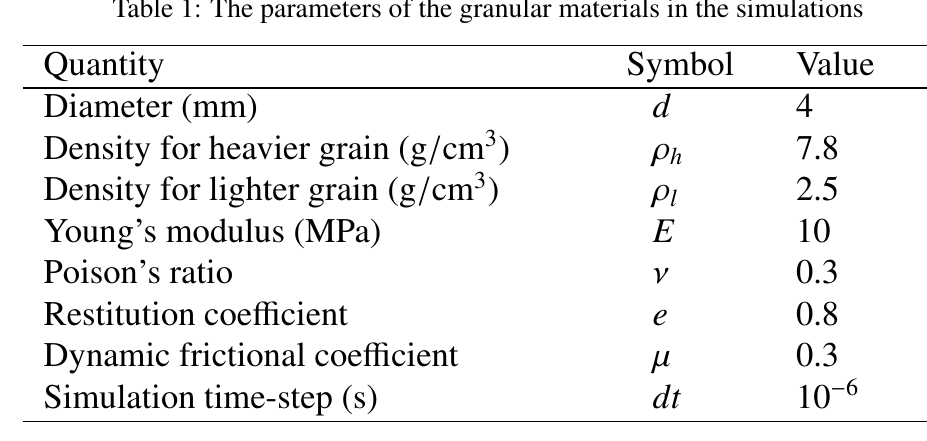} 
  \label{fig:Table1}
\end{figure}
\begin{figure*}[htbp]
\centering
\includegraphics[width=1.0\textwidth,trim=11 18 11 13,clip]{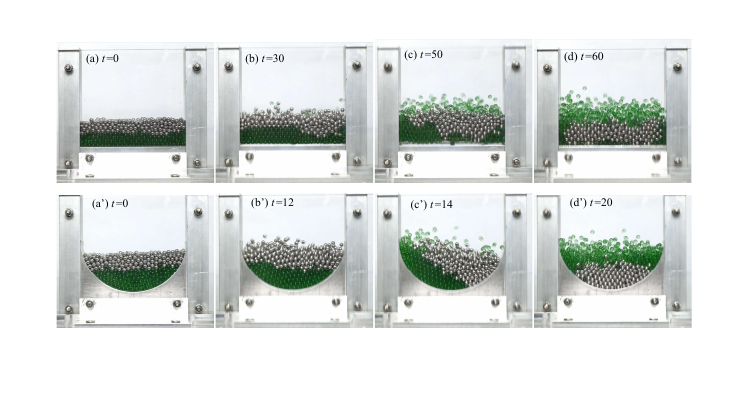}
  \caption{Experimental snapshots of the flat and circular bottom containers for $\theta=0$ in (a)(b)(c)(d) and $\theta=\pi/2$ in (a')(b')(c')(d'), respectively.}
  \label{fig:FigPositionExp}
\end{figure*}
\begin{figure*}[htbp]
\centering
\includegraphics[width=1.0\textwidth,trim=0 15 2 2,clip]{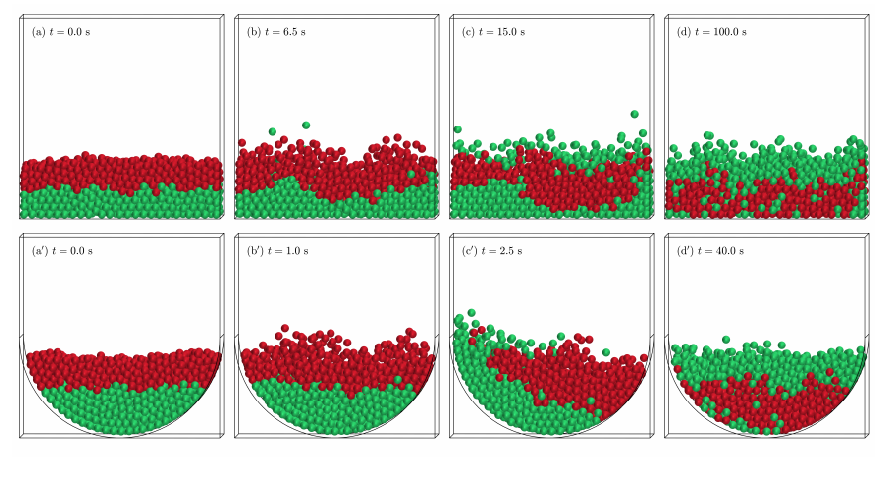} 
  \caption{Simulational snapshots of the flat and circular bottom containers for $\theta=0$ in (a)(b)(c)(d) and $\theta=\pi/2$ in (a')(b')(c')(d'), respectively.} \label{fig:FigPositionSim}
\end{figure*}
\section{RBNE-BNE segregation inversion} 
\label{SegPattern}

Fig.\ref{fig:FigPositionExp} shows two typical temporal evolutions of the experimental segregation states of binary mixtures in the flat-bottom and the circular-bottom containers, $\theta=0$ and $\pi/2$, respectively.
An initial two-layer RBNE pattern was prepared shown in Fig.\ref{fig:FigPositionExp}(a), in which glass grains (lighter grains) were first randomly poured into the container and then stainless steel grains (heavier grains) were stacked at the top layer.
The packing height of each species is about $4$ layers.
At the beginning of vibration, all grains are excited due to the vibration of the container.
Most glass and steel grains oscillate up and down as an entire ensemble, while few stainless steel grains on the surface are agitated into a gas-like state.
As the vibration continues, more stainless steel grains participate in the gas-like state, and few glass grains start to appear at the top, as shown in Fig.\ref{fig:FigPositionExp}(b).
For the vibrated container having a flat bottom, the initial packing history has a great influence on the starting location where the sinkage of heavier grains occurs.
However, a certain event is that steel grains with instantaneous local aggregation are more likely to sink compared with individual grains.
In the meantime, lighter glass grains at the bottom are pushed upward and most of them arise along the sidewalls in our experiments.
Once the sinkage of accumulated steel grains is triggered, more steel grains join the sinking process in turn, as shown in Fig.\ref{fig:FigPositionExp}(c).
The process of segregation inversion is irreversible until most heavier steel grains migrate to the bottom.
Fig.\ref{fig:FigPositionExp}(d) shows the universal final pattern, in which the heavier steel grains stay at the lower bottom and the lighter glass grains settle at the top layer.

The same experiments are conducted for a circular-bottom container, $\theta=\pi/2$, whose initial piling state is also the RBNE pattern, as shown in Fig.\ref{fig:FigPositionExp}(a').
All grains start to oscillate with the container vibration at the beginning time.
The first gasified grains are stainless steel grains on the surface, as depicted in Fig.\ref{fig:FigPositionExp}(b').
Nonetheless, compared with the segregation inversion between the RBNE and BNE states in the flat-bottom container, the starting location, where the heavier grains sink first, seems to have no close dependence on the initial packing state.
The onset of the sinkage of heavier steel grains always occurs at the either of the sidewalls.
In Fig.\ref{fig:FigPositionExp}(c'), more steel grains have accumulated near the right sidewall and glass grains are being squeezed to the left side of the container.
As this process continues, a stable BNE segregation pattern is finally reached as shown in Fig.\ref{fig:FigPositionExp}(d'), in which heavier steel grains migrate to the container bottom and glass grains rise up to the top layer.

Based on the above-mentioned experimental results in Fig.\ref{fig:FigPositionExp}, the RBNE-BNE segregation inversion in the vibrated circular-bottom container is obviously faster than that in the flat-bottom container.
This general characteristic is supported by corresponding simulations as shown in Fig.\ref{fig:FigPositionSim}.
The initial piling state is also prepared in the RBNE pattern shown in Fig.\ref{fig:FigPositionSim}(a)(a').
When the vibration starts, three segregation inversion processes from the RBNE to the BNE pattern are clearly repeated.
(I) Initial agitation process. At the beginning, under the agitation of vibration, the grains are rapidly excited in the vertical direction, especially for those on the surface, as shown in Fig.\ref{fig:FigPositionSim}(b)(b').
(II) Segregation inversion process. The initial piling history exerts a great influence on the starting location, where the heavier grains first sink down for in the flat-bottom container.
However, in the circular-bottom container, heavier grains near the sidewalls go down first.
The RBNE-BNE segregation inversion is irreversible once the aggregated heavier grains start to sink until the process is finished, as illustrated in Fig.\ref{fig:FigPositionSim}(c)(c').
(III) Steady BNE pattern process. The BNE pattern is a unique final pattern, in which the heavier and lighter grains stay at the bottom and upper layers, respectively, as shown in Fig.\ref{fig:FigPositionSim}(d)(d').

To quantify the RBNE-BNE segregation inversion, one can define a segregation factor $\chi_z$ in accordance with the center position of grains in $z$ direction, $\bar{z}_p={\frac{1}{N_p}}\sum\nolimits_{i=1}^{N_p}z_i$, where $p=h,l$ is the species of grains.
$N_p$ is the number of the $p$ species. $z_i$ is the height of grain $i$ with respect to the container bottom.
Segregation factor $\chi_z$ in the vertical direction is written as \cite{Huang2012PRE85.031305,Huang2013EPJE36.41,Hsiau2020APT31.94}:
\begin{equation}
\chi_{z}=2\frac{\bar{z}_l-\bar{z}_h}{\bar{z}_l+\bar{z}_h},
\label{eq:Segfactorz}
\end{equation}

\noindent where $\bar{z}_h$ and $\bar{z}_l$ are the center positions of the heavier and the lighter grains, respectively.
The BNE pattern occurs for positive $\chi_z$, whereas the RBNE pattern occurs for negative $\chi_z$. Similarly, to quantify the degree of segregation in $x$ direction, segregation factor $\chi_x$ is introduced as follows:
\begin{equation}
\chi_{x}=2\frac{\bar{x}_l-\bar{x}_h}{W},
\label{eq:Segfactorx}
\end{equation}

\noindent where $\bar{x}_p={\frac{1}{N_p}} \sum\nolimits_{i=1}^{N_p}x_i$ is the center position of the grains in $x$ direction. Negative $\chi_{x}$ means that lighter and heavier grains are located on the left and right sides of the container respectively, whereas positive $\chi_{x}$ means that they exchange their locations.
Apparently, large amplitudes of $\chi_{z}$ and $\chi_{x}$ suggest a relatively strong separation state in the vertical and horizontal directions, respectively.
\begin{figure}[htbp]
\centering
\includegraphics[width=0.495\textwidth,trim=5 5 5 2,clip]{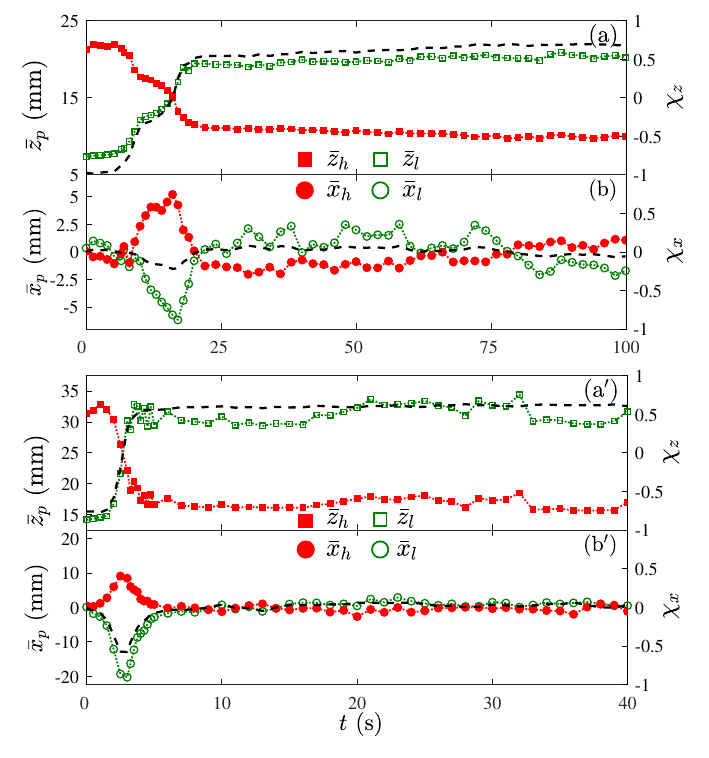} 
  \caption{Temporal evolutions of the mass-center position of each grain species for (a)(b) the flat-bottom container and (a')(b') the circular-bottom container. The squares and circles indicate the positions in $z$ and $x$, directions, respectively. The solid and open symbols are for the heavier and light grains, respectively. The dashed lines indicate the segregation factores of $\chi_z$ and $\chi_x$.}
  \label{fig:FigPositionTime}
\end{figure}

Fig.\ref{fig:FigPositionTime}(a) shows the temporal evolutions of the mass-center position of each species of grains in the container with a flat bottom, where three segregation processes are quantitatively illustrated. In the initial agitation process, $\bar{z}_h$ has a larger value, whereas $\bar{z}_l$ has a smaller value, which indicates the initial RBNE pattern.
When the segregation inversion starts, $\bar{z}_h$ and $\bar{z}_l$ have a completely reversal change.
The former decreases, and the latter increases continuously, which is consistent with the sinkage of heavier grains and the rise of lighter grains.
This characteristic is kept until the whole segregation inversion from the BNE pattern to the RBNE pattern is finished.
Both $\bar{z}_h$ and $\bar{z}_l$ reach stable values respectively, in which $\bar{z}_h$ is smaller than $\bar{z}_h$. Heavier grains stay at the low layer, and lighter grains migrate to the top layer.
In the meantime, heavier and lighter grains exchange their positions in $x$ direction. In the initial agitation process, both $\bar{x}_l$ and $\bar{x}_h$ fluctuate near zero, which means that heavier and lighter grains are uniformly distributed in $x$ direction.
Then, they start to exchange their horizontal positions of $\bar{x}_h$ and $\bar{x}_l$, where both lighter and heavier grains first move toward the left and right sides of the container respectively. Afterwards, they turn round and migrate toward the right and left sides of the container.
In the final steady state, both $\bar{x}_l$ and $\bar{x}_h$ return to near zero which means that heavier and lighter grains are mixed again in $x$ direction.

In Fig.\ref{fig:FigPositionTime}(a), three silimar processes of the RBNE-BNE segregation inversion is quantified by the temporal evolution of the segregation factor of $\chi_z$.
In $z$ direction, larger values of negative and positive $\chi_z$ indicate the initial RBNE and the final BNE patterns, respectively. In the middle range, the continuous increase in $\chi_z$ implies that segregation inversion is happening.
The temporal evolution of $\chi_x$ is plotted shown in Fig.\ref{fig:FigPositionTime}(b).
In the initial and final ranges, $\chi_x$ fluctuates around $0$, which denotes a mixed state in $x$ direction.
$\chi_x<0$ suggests that the heavier and lighter grains are located at the right and left sides of the container, respectively.

Similar simulation results of the RBNE-BNE segregation inversion are observed in the circular-bottom container, as shown in Fig.\ref{fig:FigPositionTime}(a')(b').
Three processes of the RBNE-BNE segregation inversion can be determined from the changes in the mass center of each species.
The transition time of the RBNE-BNE segregation inversion $t_{\rm{inv}}$ can be defined as the point $\chi_z=0$.
The transition times are $16.1$ and $2.66\rm{s}$ for the flat-bottom container and circular-bottom containers, respectively.
This result suggests that the latter has a higher transition speed for the RBNE-BNE segregation inversion.
\begin{figure*}[htbp]
\centering
\includegraphics[width=0.8\textwidth,trim=2 00 2 00,clip]{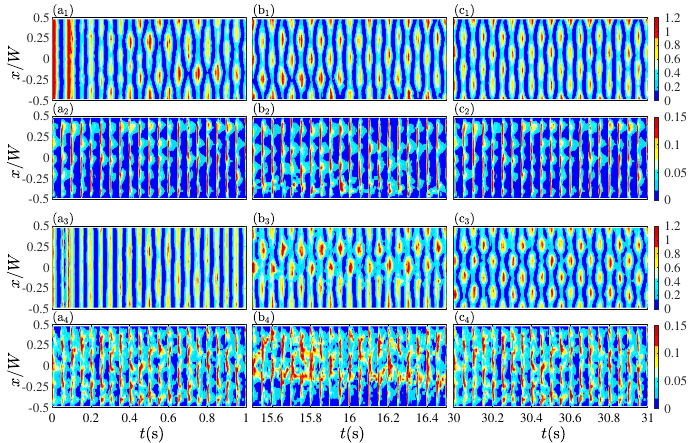} 
  \caption{Spatial-temporal evolution of granular temperature scaled by $T_0=0.5v_0^2$ in the flat-bottom container. In $\rm{(a_1),(b_1),(c_1)}$, and $\rm{(a_2),(b_2),(c_2)}$, $T_z$ and $T_x$ are plotted for heavier grains, respectively. In $\rm{(a_3),(b_3),(c_3)}$, and $\rm{(a_4),(b_4),(c_4)}$, $T_z$ and $T_x$ are plotted for lighter grains, respectively.}
\label{fig:FigTempF}
\end{figure*}
\begin{figure*}[htbp]
\centering
\includegraphics[width=0.8\textwidth,trim=2 00 2 00,clip]{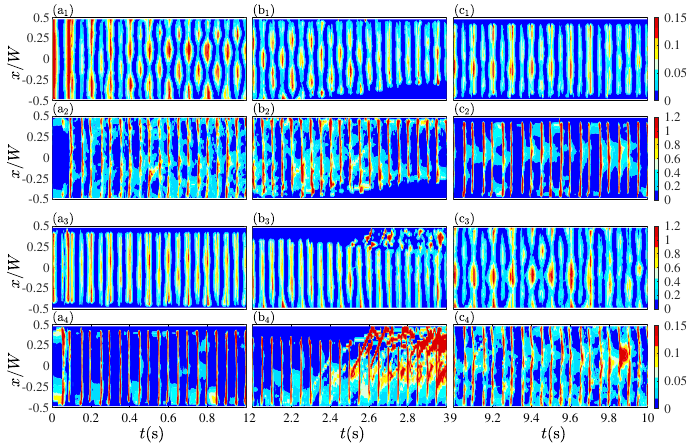} 
  \caption{Spatial-temporal evolution of granular temperature scaled by $T_0=0.5v_0^2$ in the circular-bottom container. In $\rm{(a_1),(b_1),(c_1)}$, and $\rm{(a_2),(b_2),(c_2)}$, $T_z$ and $T_x$ are plotted for heavier grains, respectively. In $\rm{(a_3),(b_3),(c_3)}$, and $\rm{(a_4),(b_4),(c_4)}$, $T_z$ and $T_x$ are plotted for lighter grains, respectively.}
\label{fig:FigTempR}
\end{figure*}
\section{Acceleration of the segregation inversion mechanism and discussion}
\label{SegMech}

In Fig.\ref{fig:FigTempF}, the spatial-temporal distribution of reduced granular temperature $T_X/T_0=v_X^2/v_0^2~(X=x,z)$ is plotted for three typical stages of the RBNE-BNE segregation inversion, in which the container bottom is flat.
For the heavier grains shown in Figs.\ref{fig:FigTempF}${\rm{(a}}_1){\rm{(a}}_2)$, both $T_z$ and $T_x$ quickly enter into a resonant state after the initial vibration period, in which a periodical standing-wave distribution of local \emph{hot} (high granular temperature) and \emph{cold} (low granular temperature) spots is clearly seen.
The positions of the hot and cold spots appear by turns at adjacent vibration periods.
This spatial-temporal dynamical characteristic is preserved through the entire process of the  RBNE-BNE segregation inversion as shown in Figs.\ref{fig:FigTempF}$({\rm{b}}_1)({\rm{b}}_2)({\rm{c}}_1)({\rm{c}}_2)$.
However, for the lighter grains, at the initial vibration stage, though $T_z$ and $T_x$ exhibit the resonant structure shown in Figs.\ref{fig:FigTempF}$\rm{(a_3)(a_4)}$, the distribution of $T_z$ shows a more uniform profile, in which the hot and cold spots are not clearly enough, compared with those in Fig.\ref{fig:FigTempF}$\rm{(a_1)}$.
When the RBNE-BNE segregation inversion is triggered, as shown in Figs.\ref{fig:FigTempF}$\rm{(b_3)(b_4)}$, the periodical distributions of hot and cold spots for $T_z$ and $T_x$ first become clearer on the right side of the container, similar to those appeared in the entire container shown in Figs.\ref{fig:FigTempF}$\rm{(b_1)(b_2)}$.
In the final steady stage, the occupied regions of hot and cold spots are expended to the entire container, as demonstrated in Figs.\ref{fig:FigTempF}$\rm{(c_3)(c_4)}$.

Fig.\ref{fig:FigTempR} shows how the spatial-temporal distribution of granular temperature develops for the circular-bottom container under the same driving conditions used in Fig.\ref{fig:FigTempF}.
As shown in Figs.\ref{fig:FigTempR}$\rm{(a_1)(a_2)}$, similar resonant states of $T_z, T_x$ are reproduced for heavier grains, in which hot and cold spots appear periodically.
For lighter grains, Fig.\ref{fig:FigTempR}$\rm{(a_3)}$ reveals a similar periodical distribution of $T_z$ in the standing-wave resonant structure.
Nevertheless, the shape of circular bottom exerts a great influence on the granular temperature distribution of $T_x$ shown in Fig.\ref{fig:FigTempR}$\rm{(a_4)}$.
The granular temperature near the sidewalls is obviously higher than that in the central regions of the container, and the periodical standing-wave distribution almost disappears.
Then, this characteristic is extended to the granular temperature of $T_x$ for heavier gains, as shown in Fig.\ref{fig:FigTempR}$\rm{(b_2)}$.
The appearance of the zero temperature region implies that the heavier and lighter grains are exchanging their positions, as depicted in $x$ direction shown in Figs.\ref{fig:FigTempR}${\rm{(b}}_1)-{\rm{(b}}_4)$.
When the RBNE-BNE segregation inversion is finished, the standing-wave structures of the temperatures of $T_z$ and $T_x$ for both heavier and lighter grains are recovered again, as shown in Figs.\ref{fig:FigTempR}${\rm{(c}}_1)-{\rm{(c}}_4)$ though the hot and cold spots are not clearly localized as those in the flat-bottom case presented in Figs.\ref{fig:FigTempF}${\rm{(c}}_1)-{\rm{(c}}_4)$.
\begin{figure}[htbp]
\centering
\includegraphics[width=0.48\textwidth,trim=0 00 0 0,clip]{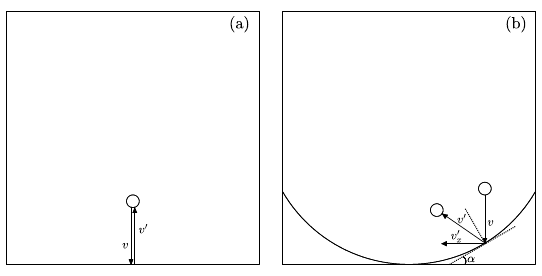}
  \caption{Sketch of the energy transfer from the vertical direction to the horizontal directions of the grain-bottom collision for (a) flat bottom and (b) circular bottom. $v$ and $v'$ denote the incident and reflected velocities respectively. $v'_x$ is for the $x$ component of the reflected velocity.}
\label{fig:CollModel}
\end{figure}

Finally, we come to the question of how the circular bottom takes an enhanced acceleration of the RBNE-BNE segregation inversion compared with that for the flat bottom.
The observed results imply that the standing-wave distribution of temperature induced by the resonance plays a dominant role in the segregation inversion.
For the development of this distribution, one precondition is the energy transfer from the vertical vibration of the container to the horizontal motion of the grains, which stems from frequent grain-grain and grain-bottom collisions.
The rotational degree of freedom is always ignored for the theoretical consideration \cite{Melo1995PRL75.3838,Huang2018PRE97.052905,Cai2020PRE101.032902,Huang2015PRE92.012202}.
As sketched in Fig.\ref{fig:CollModel}(a), no energy from vertical to horizontal directions is transferred because of the collision between the grain and the flat bottom.
The grains far from the bottom mainly obtain the agitated energy in horizontal direction by frequent grain-grain collisions.
In Fig.\ref{fig:CollModel}(b), a part of the horizontal energy is directly transferred after the collision between the grains and the circular-bottom wall.
Under the ideal collision condition, the transferred reduced kinetic energy can be written as $\Delta E_k=\frac{1}{2}{v'}_{x,i}^2$ for a single collision, where ${v'}_{x,i}=v\rm{sin}(2\alpha)$.
$v$ and $v'$ are the incident and reflected velocities respectively.
The total transferred energy is monotonically increased with the increase in the bottom angle $\theta$.
Consequently, the grains in the circular-bottom container can gain a larger kinetic energy of the horizontal component than those in the flat-bottom container.
This finding implies that more transferred energy obviously accelerates the agitation process of the granular bed.
In other words, the corresponding granular temperature $T_x$ in the circular-bottom container should be higher than that in the flat-bottom container at the same vibration conditions.
The grains around the lateral edges of the circular bottom have higher kinetic energies relative to those in the central regions.
\begin{figure}[htbp]
\centering
\includegraphics[width=0.495\textwidth,trim=110 280 110 275,clip]{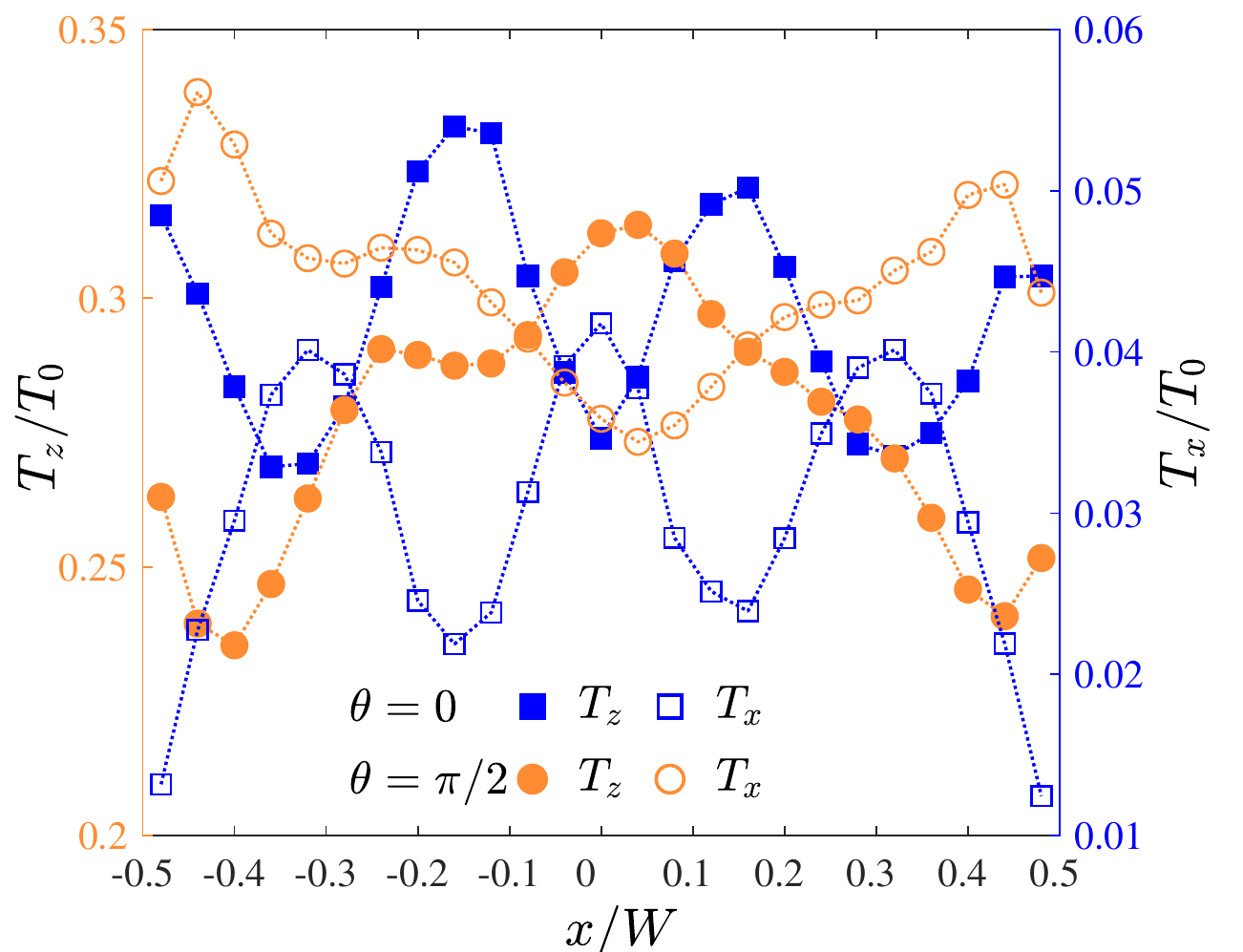}
  \caption{Reduced granular temperature as a function of $x$. The solid and open symbols denote the temperatures of $T_z$ and $T_x$, respectively. The squares and circles are for the flat-bottom container and the circular-bottom container, respectively.}
\label{fig:FigTempDistS}
\end{figure}

To explore the influence of the bottom shape on the distribution of temperature and the acceleration of the RBNE-BNE segregation inversion, simulations are performed for monodisperse grains with mass densities, i.e., $\rho=2.5,5.0$ and $7.8~\rm{g/cm^3}$.
Similar results are obtained, and only the results of $\rho=7.8$ are shown in Fig.\ref{fig:FigTempDistS} where the container is divided into $25$ subregions in $x$ direction.
A general periodical standing-wave distribution is found for both the flat-bottom and the circular-bottom containers.
The former has a relatively flat profiles with almost the same maximum temperature for both $T_z$ and $T_x$, which implies that the granular bed is agitated uniformly.
On the contrary, a higher $T_z$ and a lower $T_x$ at the lateral edges than those in the central regions appear because the circular bottom has a nonuniform energy transfer.
These characteristics are consistent with the findings from the theoretical analysis of Fig.\ref{fig:CollModel}.
The oblique grain-bottom collision due to the circular bottom results in more energy transferred; and thus, the maximum granular temperature $T_x$ happens at the lateral edges of the circular bottom.
\begin{figure}[htbp]
\centering
\includegraphics[width=0.45\textwidth,trim=135 225 150 235,clip]{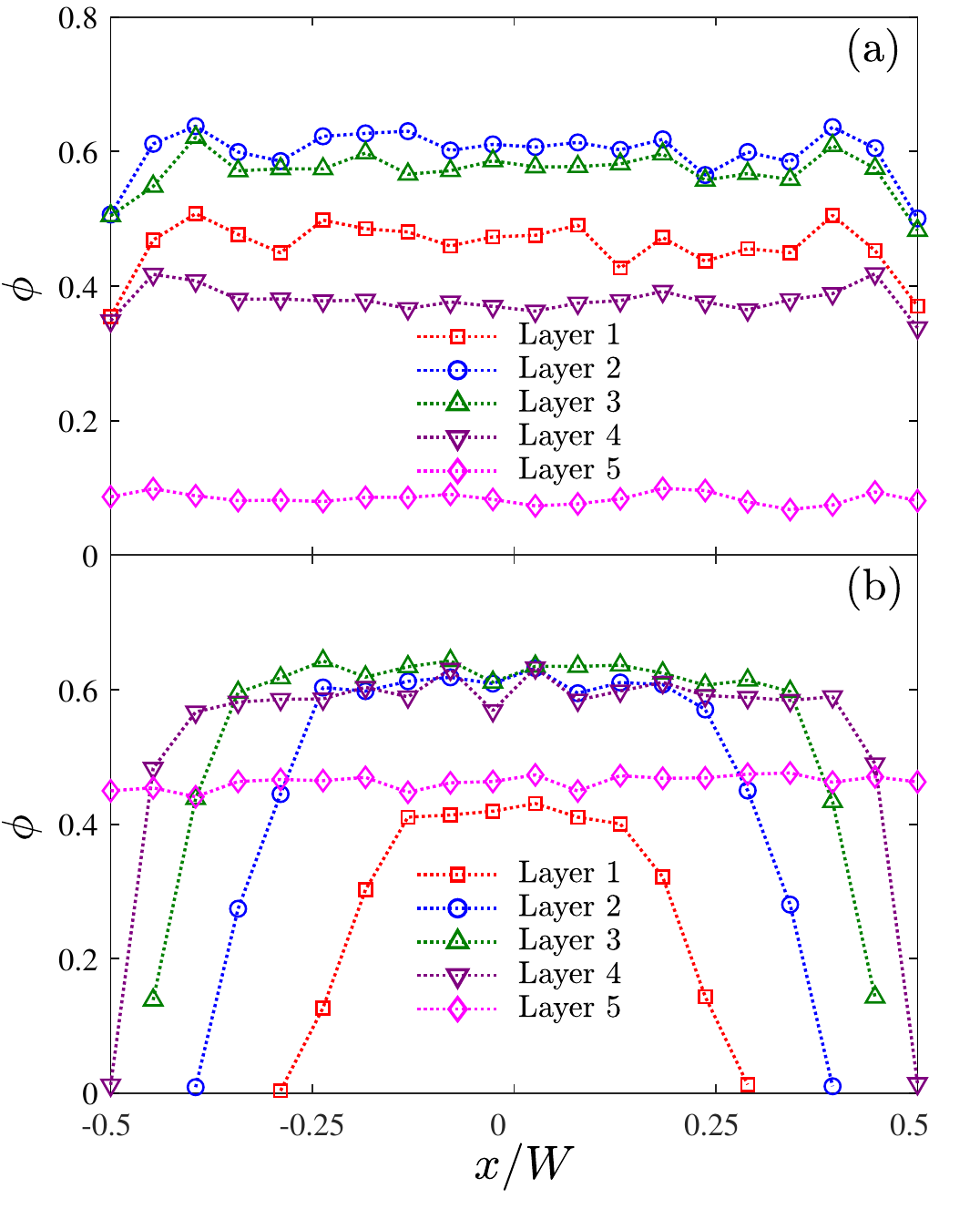}
  \caption{Packing density as a function of $x$. The squares, circles, up-triangles, down-triangles and diamonds denote layers 1,2,3,4, and 5, respectively. The dotted lines are guide for eyes.}
\label{fig:FigPackingX}
\end{figure}

Fig.\ref{fig:FigPackingX} compares the distribution of packing density for the flat-bottom and circular-bottom containers, in which the same vibration conditions as those used in Fig.\ref{fig:FigTempDistS} are considered.
The container is divided into $20\times25$ subregions in $x-z$ plane.
First, a higher packing density appears in the middle layers in $z$ direction for both the containers, which has already been noted as the granular Leidenfrost effect\cite{Eshuis2005PRL95.258001,Zhang2022PT399.117158}.
Second, all layers have the same flat profile of packing density for the flat-bottom container.
However, for the circular-bottom container, most layers hold an arch profile, in which the packing density near the lateral edges is much less than that in the central regions, except the surface layer.
\begin{figure}[htbp]
\centering
\includegraphics[width=0.495\textwidth,trim=90 280 110 275,clip]{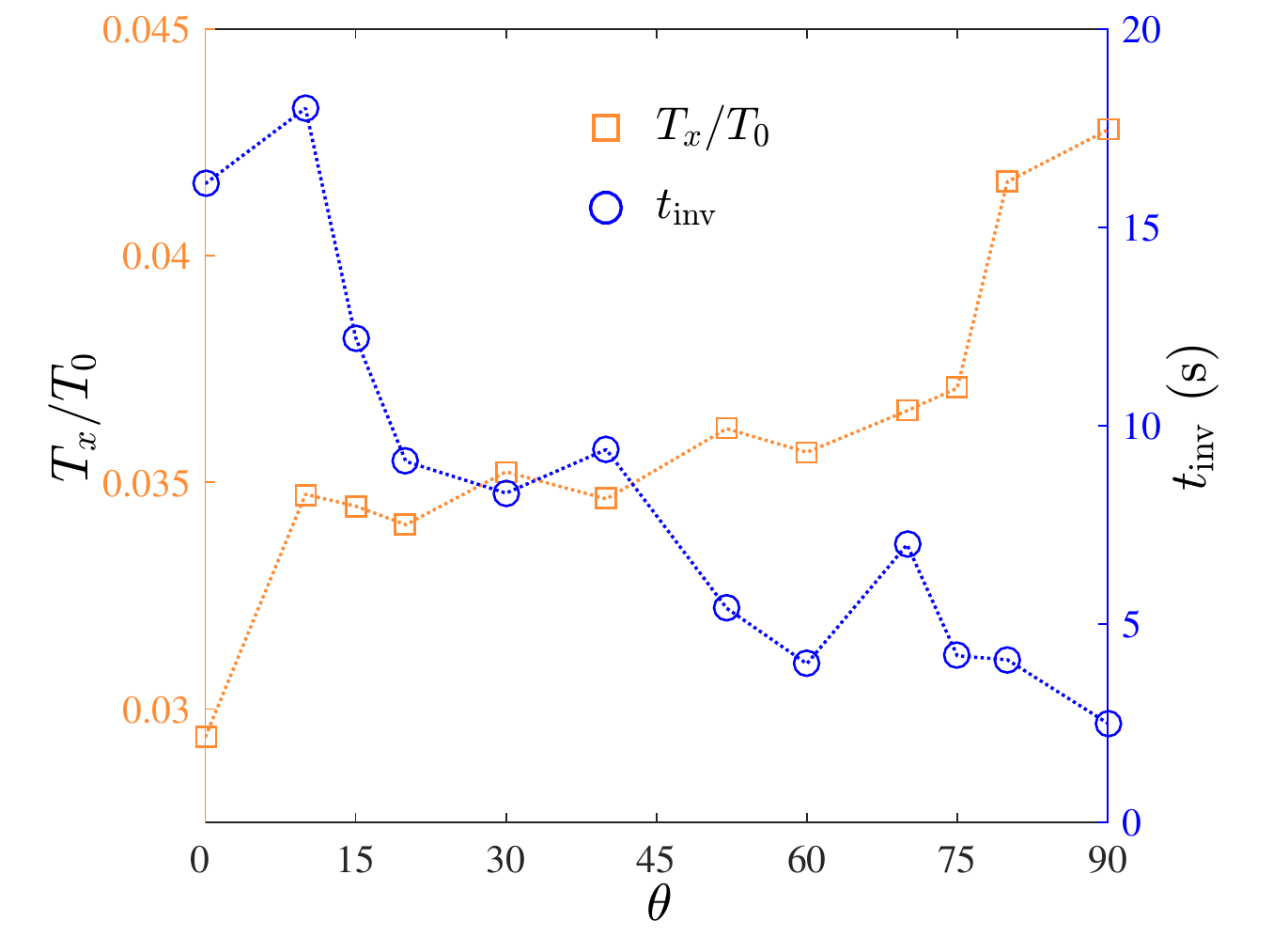}
  \caption{Granular temperature $T_x$ and transition time $t_{\rm{inv}}$ as a function of bottom angle $\theta$. The squares and circles denote $T_z$ and $t_{\rm{inv}}$, respectively.}
\label{fig:Fig10TempTime}
\end{figure}

Based on the above observations and analysis, the occurrence of the standing-wave structure of temperature induced by resonance determines the acceleration of the segregation inversion.
Owing to the discrete characteristic, the high local temperature, which means that the grains have a large momentum, could lead to the appearance of a strong collision between the grains, and a low local packing density is created.
Following the arguments on the buoyancy effect, heavier grains are more easily sink down in the local region of lower packing density.
The RBNE-BNE segregation inversion in experiments and simulations can be explained by the different energy transfer modes induced by flat bottom and circular bottom.
When the bottom is flat, the granular bed is uniformly \emph{heated} upon the vertical vibration of the bottom, which creates a flat periodical distribution of resonant states.
When the bottom becomes circular, the higher energy transfer efficiency from the $z$ direction to the $x$ direction due to the oblique grain-bottom collisions results in the fast appearance of resonant state of the granular bed, especially at the edges of the circular bottom.
Consequently, the starting location of the RBNE-BNE segregation inversion is uncertainty for flat-bottom container, whereas it first occurs at the lateral edges for the circular-bottom container.

Fig.\ref{fig:Fig10TempTime} summarizes the granular temperature $T_x$ and transition time $t_{\rm{inv}}$ as a function of bottom angle $\theta$.
As expected from the theoretical analysis, granular temperature $T_x$ increases monotonously with the increase of bottom angle $\theta$.
Furthermore, the oblique collision in the circular-bottom container induces a surge in energy transfer compared with that in the flat-bottom container.
The dependence of transition time $T_{\rm{inv}}$ on bottom angle $\theta$ also agrees well with the theoretical prediction, in which a higher granular temperature results in a faster occurrence of the RBNE-BNE segregation inversion.

\section{Conclusions}
\label{Conclusions} 

In this study, the RBNE-BNE segregation inversion of binary mixtures in the flat-bottom and the circular-bottom containers is investigated by experiments and simulations.
The starting location of sinkage of heavier grains at the top layer occurs with certain randomness in the flat-bottom container, whereas it appears first at the either of the sidewalls in the circular-bottom container.
The segregation factors in the vertical and horizontal directions are defined to quantify the degree of segregation.
The entire segregation inversion can be divided into initial heating, segregation inversion transition and steady BNE states.
A faster transition of the RBNE-BNE segregation inversion in the circular-bottom container than that in the flat-bottom container is observed.

Further simulation results present that the standing-wave resonant spots of higher and lower temperature play a dominant role in the acceleration of the RBNE-BNE segregation inversion.
In general, the heavier grains at the top layer are easier to be agitated into the standing-wave resonant state denoted by granular temperature in both vertical and horizontal directions.
For the flat-bottom container, the lighter grains gradually enter into the standing-wave resonant state and are extended to the entire container.
When the bottom is circular, both heavier and lighter grains at the lateral edges of the bottom achieve a higher granular temperature than those in the central regions of the container.
The resonant spots of higher and lower temperature still exist, though they are not clear as those observed in the flat-bottom container.

The elastic collision model of single grain shows that the oblique collision between the grain and the circular bottom holds a higher energy transfer efficiency from the vertical direction to the horizontal direction than that between the grain and the flat bottom.
The simulation results of a vibrated monodisperse granular bed show that uniform standing-wave distributions of temperature and packing density occur in the flat-bottom container.
By contrast, the granular bed has a higher temperature in the horizontal direction and a lower packing density at the lateral edges of the bottom in the circular-bottom container.
We conclude that the grain with higher granular temperature has a larger momentum, which leads to a stronger collision with the other grains.
An instantaneous lower packing density is produced, and the buoyancy effect drives the surrounding heavier grains to sink downward easily at the resonant hot spots.
A group of simulation results confirm that the granular temperature in the horizontal direction increases with the increase in bottom angle.
The occurrence of a lower packing density at the lateral edges of the bottom is favorable for faster sinkage of heavier grains happens due to the buoyancy effect.
Thus, increasing the bottom angle leads to an acceleration of the RBNE-BNE segregation inversion.
The presented results in this paper increase the understanding of the driving mechanism of the segregation of binary mixtures.
Further work is encouraged to explore the influences of grain size and vibration frequency on the resonance effect and the convection of granular beds under vertical vibration.
Related studies have potential theoretical and industrial implications in consideration of the importance of processing granular mixtures.


\begin{acknowledgments}
This work are financially supported by the National Natural Science Foundation of China (Grant No. 11574153) and the fund of No.TSXK2022D007.
\end{acknowledgments}

\end{document}